\newcommand{\HgBCO}{{$\mathrm{HgBa_{2}CuO_{4+\delta}}$ }} 
\newcommand{\eg}{\textit{e.g.}}

\documentclass[12pt]{iopart}

\usepackage{graphicx}  

\begin{document}

\title[Phonon dispersion in Hg-1201]{Phonon dispersion in the 1-layer cuprate 
$\mathbf{HgBa_{2}CuO_{4+\delta}}$}

\author{Matteo d'Astuto\dag \footnote[3]{Present address:
Physique des Milieux Condens\'es, Universit\'e Pierre et Marie Curie, 
CNRS UMR 76 02, 
F-75252 Paris C\'edex 05, France}, 
Alessandro Mirone\dag, Paola Giura\dag, Doroth\'ee Colson\ddag, 
Anne Forget\ddag~ and Michael Krisch\dag}
\address{\dag European Synchrotron Radiation Facility, 
BP 220, F-38043 Grenoble Cedex, France}
\address{\ddag Service de Physique de l'Etat Condens\'e, CEA Saclay, F-91191 
Gif-sur-Yvette Cedex, France}
\ead{matteo.dastuto@pmc.jussieu.fr}

\begin{abstract}
We investigate the low energy acoustical 
and optical modes in $\mathrm{HgBa_2CuO_{4+\delta}}$ 
using inelastic x-ray scattering (IXS). 
The experimental phonon dispersion and the dynamical structure factor 
are compared with an atomic shell model, and 
the set of the atomic potentials obtained are discussed. 
Our results are also compared with those obtained by Raman spectroscopy 
and with density-of-state data measured by 
inelastic neutron scattering. 
\end{abstract}

\pacs{74.72.Jt, 74.25.Kc, 78.70.Ck}

\maketitle 

\section{Introduction}

\begin{figure}
\vspace{18pt}
\begin{center}
\includegraphics[scale=0.4]{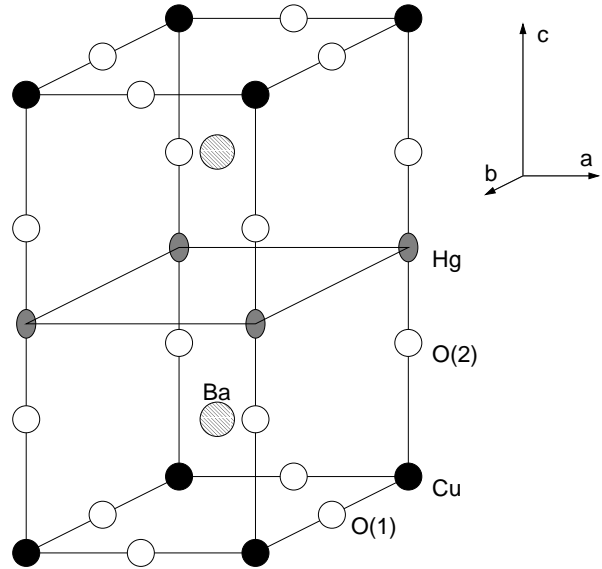} 
\end{center}
\caption{\label{crystal} Crystal structure of 
$\mathrm{HgBa_{2}CuO_{4+\delta}}$. O(1) \textit{in-plane} oxygen, 
O(2) \textit{out-of-plane} oxygen,
a = b = 3.874(1) \AA, c = 9.504(9) \AA, 
$z$O(2) = 0.2092(6) r.l.u., $z$Ba = 0.2.991(1) r.l.u. (Bertinotti \etal 1996).}
\end{figure}

The cuprate family of $\mathrm{HgBa_{2}Ca_{n-1}Cu_{n}O_{2n+2+\delta}}$ 
(or Hg-12(n-1)n) has the highest known superconducting transition temperature 
at ambient pressure, with a T$_c$ of 136K  
for the 3-layer compound 
$\mathrm{HgBa_{2}Ca_{2}Cu_{3}O_{8+\delta}}$ (Hg-1223)
(Putilin \etal 1993, Schilling \etal 1993). 
Moreover, its superconducting gap shows a peculiar symmetry, 
different from the $d_{x^2-y^2}$ symmetry 
found in the majority of the other cuprates,  
as Sacuto \etal (1997) clearly proved for 
$\mathrm{HgBa_{2}CaCu_{2}O_{6+\delta}}$ (Hg-1212). 
Within this family, $\mathrm{HgBa_{2}CuO_{4+\delta}}$ 
(Hg-1201) is perhaps even more intriguing, 
being the 1-layer hole doped cuprate with the highest superconducting 
transition temperature T$_c$ = 97 K, more than twice 
the one of the other 1-layer cuprates (25 K - 40 K). 
At the same time it presents 
one of the simplest crystal structures, a tetragonal primitive 
P/4mmm (see figure \ref{crystal}) symmetry, with very little impurities, 
no distortions and almost perfect square Cu-O$\mathrm{_2}$ planes 
(Bertinotti \etal 1997). 
For these reasons, Hg-1201 is considered as a prototype 
among the superconductor cuprates and relevant informations 
on the origin and on the detailed microscopic 
mechanism of the superconductivity are expected from the study 
of the collective excitations dynamics (electronic, crystalline, 
\eg~ phonons, and magnetic\footnote{For a full discussion 
of the superconductivity in cuprates and the role 
of the different collective excitations in these systems 
we refer the reader to the review by Orenstein and Millis (2000), 
Hussey (2002) and Moriya and Ueda (2003), 
while, for a discussion of the role of phonons in cuprates, 
to Pintschovius and Reichardt (1998) as well as  
the recent works of Anderson (2002), Bohnen \etal (2002),
Chung \etal (2003) and Bishop \etal (2003) 
and the references therein.})  
of this system.
However, such kind of study is experimentally not trivial since 
the synthesis of superconducting Hg-1201 single crystals
is possible only for volumes smaller then 0.1 mm$^3$ 
(see Bertinotti \etal 1997). 
Such a size is insufficient for inelastic neutron scattering (INS), 
that is the standard technique used to measure the dispersion 
of phonons. 
For this reason, phonon data on this system were, up to now, 
limited to the ones obtained by INS performed on powders, 
which gives a measure of the Phonon Density Of State 
(PDOS) (Renker \etal 1996) and by Raman spectroscopy  
(Hur \etal 1993, 
Krantz \etal 1994, Lee \etal 1994, Ren \etal 1994, 
Yang \etal 1995,
Zhou \etal 1996a, 1996b, 
Poulakis \etal 1999 and 
Cai \etal 2001). These experimental results have been analysed using 
a shell model calculation by Renker \etal (1996) and Stachiotti \etal (1995), 
who have also determined the zone centre frequencies from frozen-phonon 
first principle calculations. 

Nowadays, the study of the phonon dispersion in 
sub-millimeter size cuprate crystals, is possible by means of 
inelastic x-ray scattering (IXS), as it has been shown  
by d'Astuto \etal (2002, 2003), 
and, very recently, by Fukuda \etal (2003).
Nevertheless, the measurements of phonons by IXS still represent 
a difficult task, since
the relevant scattering cross section, $\propto f(Q)^2 \propto Z^2$, 
is due to the oxygen motion, while the 
relevant penetration depth, $\propto \frac{1}{Z^4}$, 
is dominated by the high Z ions 
(neodymium in d'Astuto \etal (2002, 2003) 
or mercury and barium in the present work), 
a fact that strongly reduces the scattering volume. 
Moreover, the tails of the elastic   
as well as the low energy phonon lines,
to which all the atoms contribute, rise the background under the 
weaker high energy phonons.   
For further detail see d'Astuto \etal (2003). 
Nevertheless, despite these experimental difficulties, 
Hg-1201 still remains, thanks to its simple structure, one of the most
suitable systems for the understanding of the phonon dispersion 
and consequently of the effects of the electron-phonon coupling.\\

In this first investigation, we focus the analysis on the Hg-1201 
low energy acoustical and optical modes, 
both in longitudinal and transverse polarisation. A particular 
attention is dedicated to the calculation of the dispersion of 
the transverse acoustic ones, in order to establish 
an accurate theoretical model able to 
predict the energy position and the intensity of the phonons 
along the dispersion curve.
A careful estimation
of the dynamical structure factor is, indeed, fundamental 
to select the optimal Brillouin zone which maximise 
the intensity of the high energy modes that scatter weakly the X-rays, 
as well as to assign the correct character to the observed branches.

\section{Experiment} \label{exp}

\subsection{Samples}

\begin{figure}
\vspace{18pt}
\begin{center}
\includegraphics[scale=0.4]{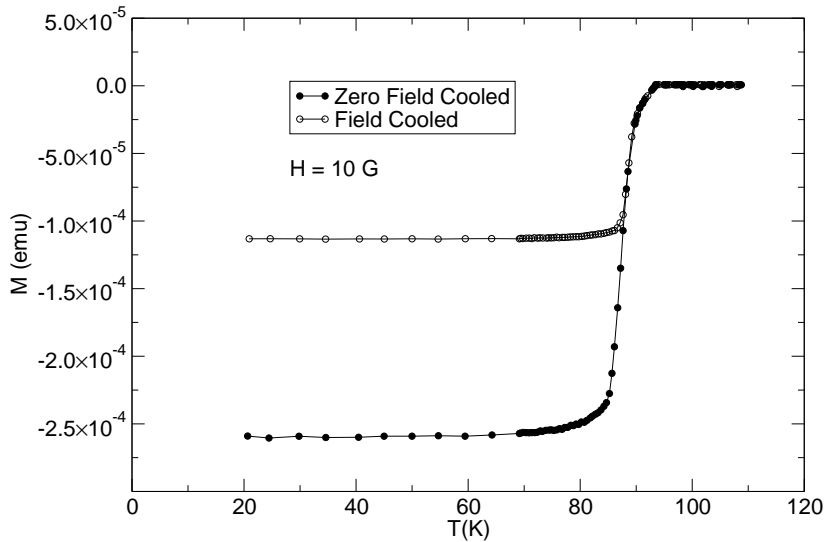} 
\end{center}
\caption{\label{squid} Magnetisation as a function of temperature for 
the sample used in the present experiment.}
\end{figure}

The single crystals of Hg-1201 were successfully grown by the flux method. 
The procedure for crystal growth will be described elsewhere 
(Colson \etal to be published).
The sample size is about 1$\times$1$\times$0.1mm, with the shorter size 
along the c axis. 
The sample is as-grown, with no annealing process, 
in order to have the best crystalline quality: 
typical mosaic spread around the a-direction  
is of about 0.03$^{\circ}$.
As the sample is as-grown, the superconducting transition is not sharp, 
but spread out over a temperature range of $\Delta$T $\simeq$  5 K 
(90\%T$_c$-10\%T$_c$), as shown in figure \ref{squid}, with a nearly optimal 
T$_c$ onset of about 94 K. 

\subsection{Inelastic X-ray scattering}
 
The IXS measurements were carried out on the undulator 
beam-line ID28 at the European Synchrotron Radiation Facility, Grenoble. 
The incident beam is monochromatized by a perfect plane Si-crystal 
(Verbeni \etal 1996), working in extreme backscattering geometry 
at the (9,9,9) reflection (17794 eV), 
with $\approx$ 0.7 THz resolution and a  wavelength $\lambda=$0.6968 \AA, 
and at the (8,8,8) reflection (15817 eV) with
$\approx$ 1.3 THz resolution, high flux and a wavelength $\lambda=$0.7839 \AA.
The monochromatic beam is focused onto the sample position 
by a toroidal mirror in a 250$\times$90 $\mu$m$^{2}$ spot. 
The scattered photons are analysed by a bench of five spherically-bent 
high-resolution Si analysers (Masciovecchio \etal 1996a, 1996b), 
placed on a 7~m long horizontal arm. 
The analysers are held one next to the other with a constant 
angular offset of about 1.5$^{\circ}$ 
and operate in backscattering geometry at the same reflection 
order as the monochromator. 
The sample was kept at ambient temperature in a vacuum chamber in order to 
prevent scattering from air and oxidation of the specimens. 
The sample chamber is held on a standard diffraction goniometer.
The energy ($h\nu$) scans are performed by varying the monochromator 
temperature while keeping the analyser crystals at a fixed temperature. 
Further details of the technique can be found in the reviews 
of Sette \etal (1996) and Burkel (2000). 
Krisch \etal (2002) give a detailed description of the set-up 
for the ID28 spectrometer at ESRF, as used in the present work.

IXS scans were taken at Q=G+q points of the reciprocal lattice, 
where G is the zone centre vector, and q is the reduced vector 
which corresponds to the phonon propagation vector.
Longitudinal or (\textit{quasi-})transverse 
phonon polarisation can be selected according to the relative direction 
of q and Q: parallel for longitudinal scans 
or (nearly) perpendicular in the second case. 

\subsection{Data analysis}

\begin{figure}
\vspace{18pt}
\begin{center}
\includegraphics[scale=0.5]{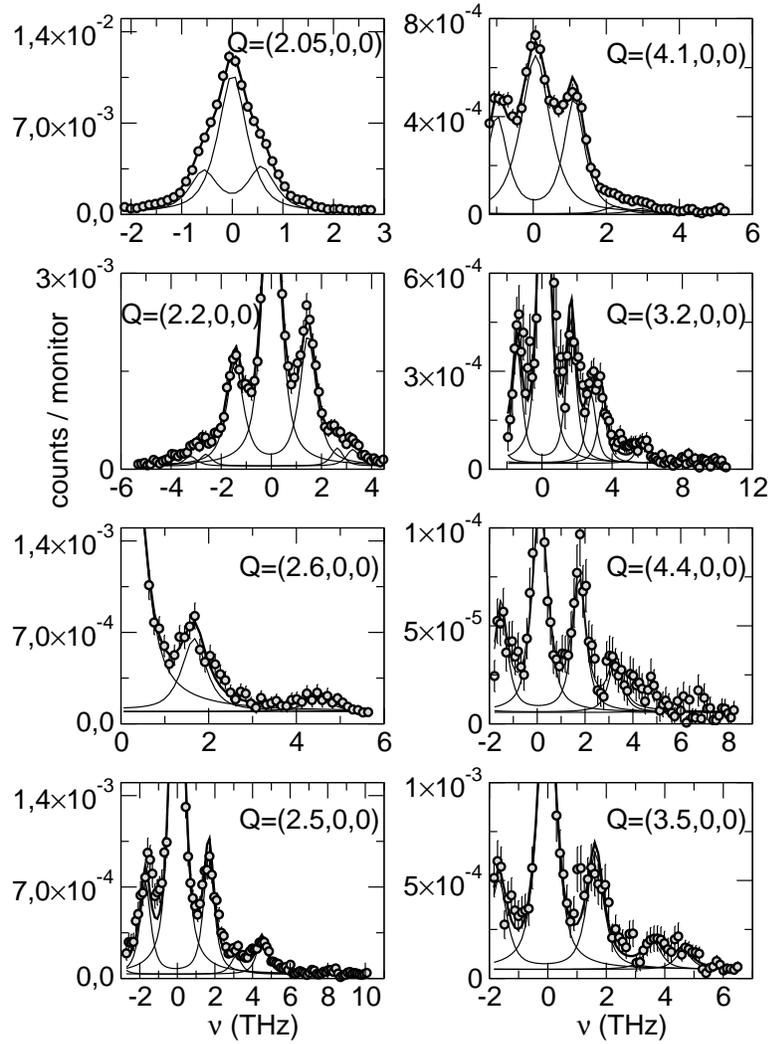} 
\end{center}
\caption{\label{qscans} Typical inelastic X-ray scattering (IXS) scans 
of \HgBCO at different scattering vectors, 
with $\sim$ 0.7 THz frequency resolution, 
normalised to the incident photon flux on an arbitrary scale, 
and lines correspond to the best fit using the model as explained 
in section \ref{exp}. 
All data have been collected at room temperature in longitudinal 
configuration (see text) parallel to the a* axes, 
with Q=G+q=(H,0,0)+($\xi$,0,0). 

\noindent
Left side: scans in the third Brillouin zone, 
with H=2 and $\xi$ between 0.05 and 0.5. 

\noindent
Right side: scans at higher Q, in the fourth and fifth zone, with H=3 and 4.
In the second and the fourth row of the figure, each pair of 
graphs correspond to the same reduced q vector, with $\xi$=0.2 and 0.5.
}
\end{figure}

The energy scans are fitted using a sum of Pseudo-Voigt functions:

\begin{equation}
I\left (
(1-\eta) \frac{\Gamma^2}{(\epsilon-\epsilon_0)^2+\Gamma^2} + \eta  
\exp \left (-\frac{(\ln 2)(\epsilon-\epsilon_0)^2}{\Gamma^2} \right )
\right ) 
\end{equation}

\noindent
where $\epsilon=\hbar\omega=h\nu$ is the energy, 
for both the elastic and 
the inelastic contributions. 
The half-width-half-maximum (HWHM) $\Gamma$ is fixed to the value obtained 
from a fit of the elastic line due to diffuse scattering.
The instrumental line-shape parameter $\eta$ was taken from a fit 
to IXS data of Nitrogen (Cunsolo \etal 2003) at 
Q$\approx$1.9 \AA$^{-1}$, P=0.25 bar and T=66.4 K, 
where the diffusion is purely elastic as established  
by Carneiro and McTague (1975). 
The line shape described above is in excellent agreement with the instrumental 
line-shape as can be seen from the elastic signal fit 
in the energy scans of figure \ref{qscans}. 
The position $\epsilon_0$ and intensity $I$ of this model 
are fitted to the \textit{IXS} signal from 
the phonons of the \HgBCO crystals 
using a $\chi^2$-minimisation routine (James and Roos 1975), 
with the condition that the detailed balance between Stokes and anti-Stokes 
excitations is fulfilled. A constant background, coming essentially from 
electronic noise, is added.

\subsection{Numerical modelling}

Phonon mode frequencies and related atomic displacements 
where obtained from the diagonalisation of the dynamical 
matrix of a classical shell model with atomic potentials, 
including screened Coulomb interactions. The matrix was built and 
diagonalised using the \textit{OpenPhonon} 
code (Mirone 2003, Mirone and d'Astuto 2003), 
as in d'Astuto \etal (2002, 2003). 

The PDOS is constructed as a 200 points histogram, 
whose frequency coordinates range from zero up to the maximum 
eigen-frequency of the system. The histogram bins are filled calculating 
the eigen-frequencies over  the Brillouin zone 
on a 11$\times$11$\times$11 grid. 
Symmetries are used to reduce the number of necessary calculations.
To get the partial DOS for a given site group, 
the contributions are weighted when added on a particular bin. 
The weight is calculated as the modulus square 
of the eigenvector projection over the considered 
degrees of freedom (Mirone 2003, Mirone and d'Astuto 2003).

\section{Results}\label{results}

\begin{figure}
\begin{center}
\includegraphics[scale=0.4]{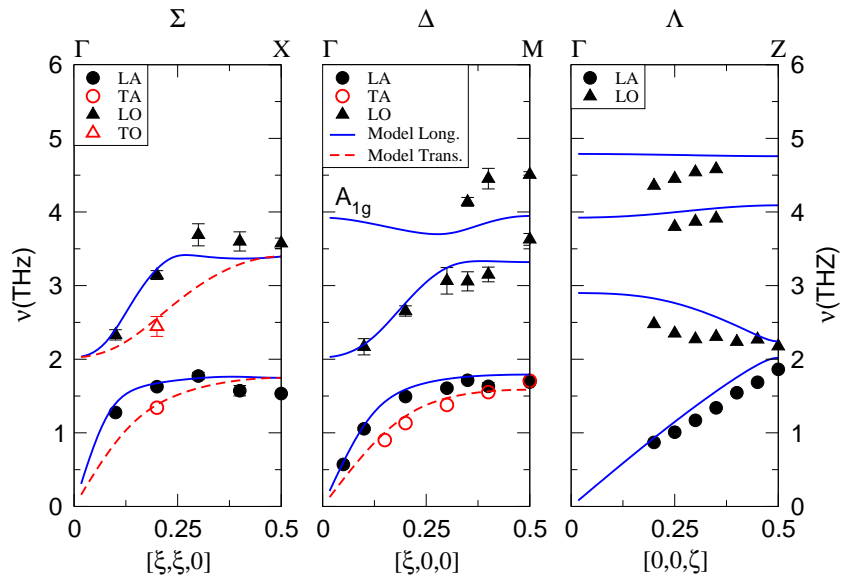} 
\end{center}
\caption{\label{disp} Phonon dispersion in \HgBCO, along the main 
symmetry directions with the reduced vector \textit{in-plane} ([$\xi$,0,0] and 
[$\xi$,$\xi$,0]), and perpendicular to the a*b* plane [0,0,$\zeta$]. 
Symbols represent the experimental frequencies determined from the IXS 
data as described in the text, continuous lines correspond to the calculate 
dispersion (see section \ref{results} and table \ref{pot}). }
\end{figure}

Figure \ref{qscans} reports some representative IXS scans in Hg-1201. 
Each graph corresponds to a different 
scattering vector, in two different Brillouin zones. 
Every scan shows a central peak, at zero frequency, 
due to the elastic diffuse scattering, 
and one or more pairs of peaks due to the inelastic scattering from phonons,
corresponding to the Stokes and anti-Stokes components of each mode.
Note that some scattering vectors correspond to the same 
reduced reciprocal lattice vector in two different Brillouin zones, 
showing the variation in intensity due to the change in the 
dynamical structure factor.  
The inelastic lines with the greater intensity are usually well resolved,  
with the exception of the scan at Q=(2.05,0,0), where the inelastic 
contribution is too close to the elastic one, resulting in a 
pair of shoulders on both sides of the elastic line.

The variation of the frequencies
of the phonon modes with 
the reduced scattering vector q, 
obtained by the IXS scans as described above, 
are shown in figure \ref{disp}, for the main symmetry direction 
\textit{in-plane} ($\Sigma$ and $\Delta$) and along c* ($\Lambda$). 
The two highest optic branches along $\Lambda$ (right panel of figure 
\ref{disp}), whose energy separation is less 
than the experimental resolution, have been measured in 
separate scans performed in two different Brillouin zones, 
according to the calculated intensities. 
Here we also show the calculated dispersion with the empiric model 
previously described, obtained using the set of atomic potentials 
reported in table \ref{pot}.
Note that a degeneracy of the \textit{in-plane} longitudinal and 
transverse modes is calculated at the $\Sigma$
zone boundary, as already 
reported by Stachiotti \etal (1995). 
Within the experimental accuracy, a similar degeneracy 
appear in the data at the $\Delta$ zone boundary.  

 \begin{table}
      \centering
      \begin{tabular}{||c|c|c|c||}

        \hline

           &  &  &  \\
        Site($\kappa$) & Z$_\kappa(e)$ & Y$_\kappa(e)$ 
        & $K_\kappa$(Nm$^{-1}$) \\
           &  &  & \\
        \hline

        Hg & 1.22 & 2 & 700 \\

        Ba & 1.69 & 3 & 700  \\

        Cu & 1.64 & 3 & 2000  \\

        O(1,2) & -1.56 & -3 & 1800 \\

           &  &  &  \\
        \hline
           &  &  &  \\
         Interaction ($\kappa-\kappa'$)  & $V_{BM}^{\circ}$(eV) 
         & $r_{\kappa, \kappa'}$(\AA) &  \\
           &  &  & \\

        \hline

         Hg-Ba & 2000 & 0.350 &  \\

         Ba-Cu & 2000 &  0.335 & \\

         Hg-O(1) & 2000 & 0.280 & \\

         Hg-O(2) & 2000 & 0.280 & $F_L=F_T$ = 40 (Nm$^{-1}$)  \\

         Ba-O(1) & 2500 & 0.315 & \\

         Ba-O(2) & 2500 & 0.325 &  \\

         Cu-O(1) & 3950 & 0.228 & \\

         Cu-O(2) & 460 & 0.353 & \\

         O(i)-O(j) i,j=1,2 & 2000 & 0.284 & 
$V_{VW}^{\circ}$ = -100 (eV \AA$^6$)  \\
           &  &  &  \\
         \hline

     \end{tabular}
\centering
\caption{\label{pot} Atomic Born von-Karman potential parameter for the 
calculation of the phonons dispersion (figure \ref{disp}, bottom panel) 
and density of state (figure \ref{dos}, bottom panel). 
For the definition of the potentials see Chaplot \etal (1995), 
and Mirone and d'Astuto (2003).}
 \end{table}  

\begin{figure}
\begin{center}
\vspace{12pt} 
\includegraphics[scale=0.5]{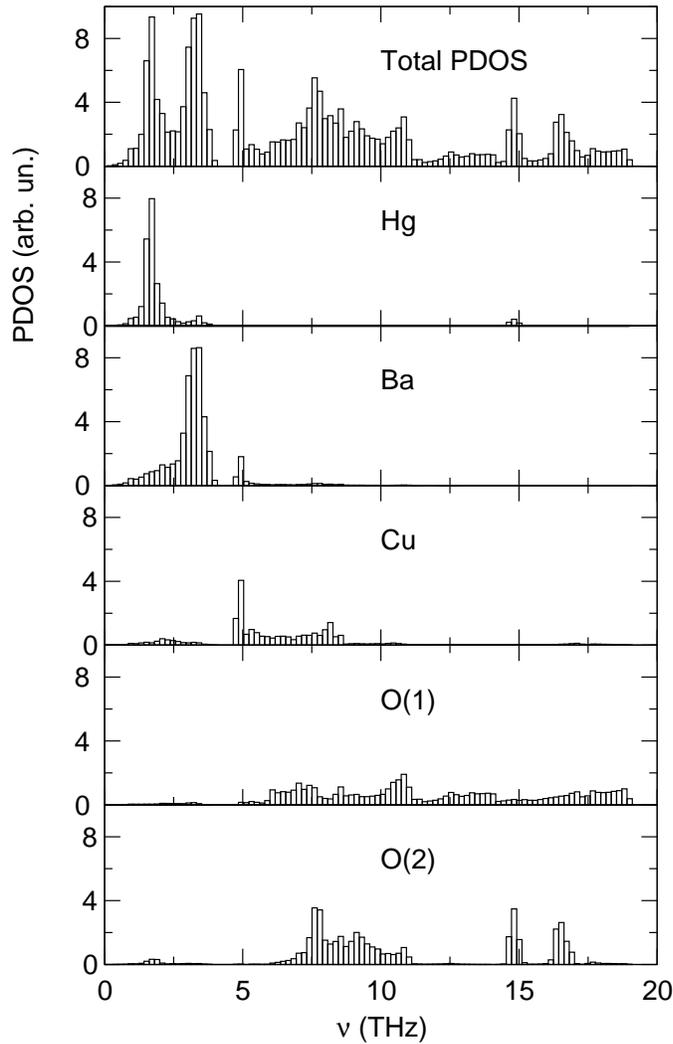} 
\end{center}
\vspace{1pt} 
\caption{\label{dos} Bar chart of the calculated PDOS.} 
\end{figure}

Most of the parameters of table \ref{pot} 
are in agreement with the values reported 
by Renker \etal (1996) and Chaplot \etal (1995). 
Nevertheless some discrepancies have to be mentioned: 
the Born - Mayer potentials for the Ba-O(1) and Ba-O(2) bonds
are different from the Ba-O potential in Chaplot \etal (1995),
with the first radius $r_{\kappa, \kappa'}$ slightly shorter and the second 
slightly larger, according to Renker \etal (1996). 
Using a mercury ion polarisability of 700 Nm$\mathrm{^{-1}}$ 
\footnote{Note that this polarisability has 
the same value given by Chaplot \etal (1995) for the barium ion}, and  
a slightly stiffer additional force constant along the Hg-O(2) 
bond, if compared to the one of Renker \etal (1996), 
we found a good agreement with our experimental dispersion. 
The histogram of the PDOS calculated 
with the parameters of table \ref{pot} 
is shown in figure \ref{dos}.

 \begin{table}
      \centering
      \begin{tabular}{||c|c|c|c||}

        \hline

        &   &  &  \\
 Mode & Exp. (cm$^{-1}$) &   Exp. (THz) &  Shell Model (THz) \\
        &   &  &  \\
        \hline
    &   &  & \\
 A$\mathrm{_{1g}}$ O(2) & 592 & 17.75 & 17.23 \\
    &   &  & \\           

 E$\mathrm{_{g}}$ O(2) & 165 &  4.95 &  7.88 \\
    &   &  & \\           

 A$\mathrm{_{1g}}$ Ba & 161  &  4.83 &  3.92  \\
    &   &  & \\           

 E$\mathrm{_{g}}$ Ba & 76  &  2.28 &  2.84 \\
           
     &   &  & \\
        \hline

     \end{tabular}
\centering
\caption{\label{raman} Comparison between 
the experimental Raman shift (Zhou \etal 1996b) 
and the Shell model (table \ref{pot}) calculation results 
for some Raman active mode. 
For a graphical representation and a symmetry analysis 
of the modes see Stachiotti \etal (1995) and Zhou \etal (1996b). }
\end{table}  

In table \ref{raman} the calculated Raman shift frequencies for 
the four Raman active modes together with the experimental ones as found by 
Zhou \etal (1996b) are reported 
(other Raman measurements, as Krantz \etal (1994), Lee \etal (1994),
Poulakis \etal (1999) and Cai \etal (2001), give similar results).
Note that the discrepancy between the calculated A$\mathrm{_{1g}}$ Ba 
mode frequency and the experimental Raman value, of about 0.9 THz, 
is also present if considering our IXS data, as reported in the central panel 
of figure \ref{disp}, although the difference in our case is only  
$\sim$ 0.5 THz.

\begin{figure}
\vspace{18pt}
\begin{center}
\includegraphics[scale=0.4]{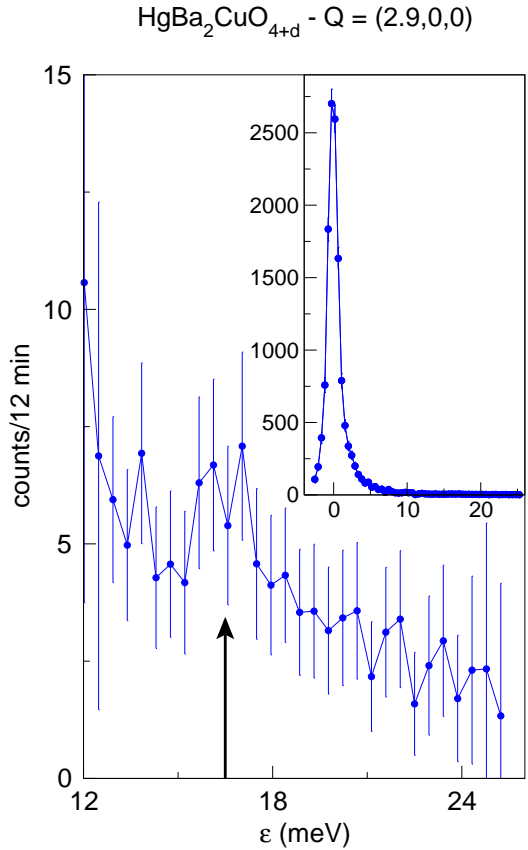} 
\end{center}
\caption{\label{hescan} Scan at Q=(2.9,0,0), with a resolution of 
1.3 THz, using the (888) monochromator reflection.}
\end{figure}

Finally, by looking at figure \ref{hescan}, in which we report an 
energy scan performed with the low resolution set up, 
we note the presence of a weak signal at high frequency. 
This signal can be assigned to the in plane oxygen vibrations.

\section{Discussion}

The main problem arising from the model calculation 
of this system using atomic potentials 
is that the transverse acoustic \textit{in-plane} modes
are not stable. 
In order to remove this instability it is necessary to add  
a force constant, which makes the bond between 
the mercury atom and the apical oxygen 
O(2) less soft against both longitudinal and transverse strains. 
This because, as pointed out by Pintschovius and Reichardt (1998),
open structures can not be stabilised by simple two body ionic 
forces, but need the rigidity of covalent bonds. 
Therefore we have optimised this force constant parameter in order 
to leave the others atomic potential compatible with the common 
potential model established by Chaplot and co-workers (1995). 
We stress here that we have not performed a fit of the experimental 
frequencies, given the limited amount of data. 
In the future, a refined model will be given on the base of a more complete 
set of measurement, including at least some of the high frequency modes.
For the present purpose we estimate that this accuracy is sufficient.

Concerning the discrepancy with the Raman data,
it has to be noted that: first Zhou and co-workers (1996b) 
express some doubts on the assignment of the signal at 4.83 THz 
and at 4.95 THz to the two modes A$\mathrm{_{1g}}$ Ba 
and E$\mathrm{_{g}}$ O(2) respectively;
second, no direct comparison with the IXS data is possible, 
since spectra at the zone centre are, in most cases, 
dominated by strong elastic scattering,
masking the weak signal from the optical branches.
Therefore we can only compare the model 
with IXS results in all the Brillouin zone but close to $\Gamma$, 
and with Raman results at the zone centre. 
Further investigation are maybe necessary, 
as suggested by Zhou \etal (1996b), 
for the correct assignment of the zone centre frequencies. 

\section{Conclusion}

We have presented the first measurements 
of the phonon dispersion in superconducting Hg-1201. 
This is an important starting point 
for further investigations aiming in detecting possible anomalies which can 
be related to an electron-phonon coupling effect, 
especially the one concerning 
the high frequency longitudinal optic mode due to oxygen vibration, 
as previously observed in other cuprate systems (Fukuda \etal 2003, 
Chung \etal 2003, d'Astuto \etal 2002, 2003 and 
Pintschovius and Reichardt 1998 and references therein). 
This search is particularly important in the system Hg-1201 as it can 
be considered as the simplest and most perfect prototype of cuprate 
superconductors. Further efforts have to be made on the experimental side, 
using IXS technique and Raman spectroscopy, 
which will greatly benefit from the present reference data and which 
can be guided by the above proposed model. The spectrum in 
figure \ref{hescan} is very promising in that sense, 
also taking into account the recent improvement 
in the spectrometer efficiency and in the sample growth.

Possible \textit{ab-initio} determinations of the phonon dispersion can also 
be very useful for this purpose, and can be previously checked 
against the present set of data.

\ack
We acknowledge G. Monaco for useful discussions and help during 
the experiments.
The authors are grateful to D. Gambetti, C. Henriquet and R. Verbeni 
for technical help and to J.-L. Hodeau for his 
help in the crystal orientation.
This work has been supported by the European Synchrotron Radiation 
Facility (Experiment HE-1369).

\section*{References}

\begin{harvard}

\item[] Anderson P W 2002 \textit{Preprint} cond-mat/0201429

\item[] Bertinotti A, 
Colson D, 
Marucco J-F, 
Viallet V, 
Le Bras G, 
Fruchter L, Marcenat C, Carrington A 
and Hammann J 
1997 
\textit{ Single Crystals of Mercury Based Cuprates: 
Growth, Structure and Physical, Properties}, in  
\textit{Studies of High Temperature Superconductors - 
Hg-Based High Tc Superconductors: Part-I} Ed. Narlikar A,
vol.~ 23 (Nova Science Publisher, New York) p~ 27-85

\item[] Bertinotti A, Viallet V, Colson D, Marucco J-F, Hammann J, 
Forget A and Le Bras G 1996 
\textit{Physica C} \textbf{268} 257-265

\item[] Bishop A R, Mihailovic D and de Leon J M 2003 
\JPCM \textbf{15} L169-L175

\item[] Bohnen K-P Heid R and Krauss M 2002 \textit{Preprint} cond-mat/0211084

\item[] Burkel E 2000
\textit{Rep. Prog. Phys.} \textbf{63} 171-232

\item[] Cai Q, Chandresekhar M, Chandrasekhar H R, Venkateswaran U, 
Liou S H and Li R 2001
\textit{Solid State Comm.} \textbf{117} 685-690

\item[] Carneiro K and  McTague J P 1975 \PR A {\bf 11} 1744

\item[] Chaplot S L, Reichardt W, Pintschovius L and Pika N 1995 
\textit{Phys. Rev. B} \textbf{52} 7230-7242

\item[] Chung J-H, Egami T, McQueeney R J, Yethiraj M, Arai M, Yokoo T, 
Petrov Y, Mook H A, Endoh Y, Tajima S, Frost C and Dogan F 2003 
\textit{Phys. Rev. B} \textbf{67} 014517-9 

\item[] Colson D \etal, \textit{to be published}

\item[] Cunsolo A Monaco G Nardone M Pratesi G and Verbeni R
2003 \PR B {\bf 67} 024507

\item[] d'Astuto M, Mang P K, Giura P, Shukla A, Ghigna P, Mirone A, 
Braden M, Greven M, Krisch M and
Sette F 2002 \PRL {\bf 88} 167002 

\item[] d'Astuto M, Mang P K, Giura P, Shukla A, Mirone A, Krisch M, Sette F, 
Ghigna P, Braden M and Greven M 2003 
\textit{Int. J. Mod. Phys. B} \textbf{17} 484-492 

\item[] Fukuda T, Mizuki J, Ikeuchi K, Fujita M, Yamada K, Baron A Q R, 
Tsutsui S, Tanaka Y and Endoh Y 2003 \textit{Preprint} cond-mat/0306190

\item[] Hur N H, Lee H-G, Park J-H, Shin H-S and Yang I-S 1993 
\textit{Physica C} \textbf{218} 365-368

\item[] Hussey N E 2002 \textit{Advances in Physics} \textbf{51} 1685-1771

\item[] James F and Roos M 1975 
\textit{Comput, Phys. Comm.} \textbf{10} 343-367

James F and Roos \textit{MINUIT - 
Function Minimization and Error Analysis - Reference Manual} 
CERN D506 (Long Writeup)
(Computing and Networks Division, CERN, Geneva)
  
available on line http://cernlib.web.cern.ch/cernlib/

\item[] Krantz M, Thomson C, Mattausch H and Cardona M 1994 
\textit{Phys. Rev. B} \textbf{50} 1165-1170

\item[] Krisch M, Brand R A, Chernikov M and Ott H R 2002 
\textit{Phys. Rev. B} \textbf{65} 134201-7 

\item[] Lee H G, Shin H S, Yang I S, Yu J J and Hur N H 1994
\textit{Physica C} \textbf{233} 35-39

\item[] Masciovecchio C, Bergmann U, Krisch M, Ruocco G, Sette F 
and Verbeni R 1996 \NIM \textbf{B 111} 181-186 

\item[] Masciovecchio C, Bergmann U, Krisch M, Ruocco G, Sette F
and Verbeni R 1996 \NIM \textbf{B 117} 339-340 

\item[] Mirone A 2003 \textit{Open Phonon Code Source, version 3} 
available on http://www.esrf.fr/computing/scientific/ 

\item[] Mirone A and d'Astuto M 2003 \textit{Open Phonon Manual, version 3} 
available on http://www.esrf.fr/computing/scientific/

\item[] Moriya T and Ueda K 2003 
\textit{Rep. Prog. Phys.} \textbf{66} 1299-1341

\item[] Orenstein J and Millis A J 2000 \textit{Science} \textbf{288} 468-474 

\item[] Pintschovius L and Reichardt W 1998  
\textit{Neutron Scattering in Layered Copper-Oxide Superconductors}, 
ed. A. Furrer, 
Physics and Chemistry of Materials with Low-Dimensional Structures 
vol. ~20 (Kluwer Academic Publishers, Dordrecht) p~ 165

\item[] Poulakis N, Lampakis D, Liarokapis E, Yoshikawa A, 
Shimoyama J, Kishio K, Peacock G B, Hodges J P,  Gameson L, 
Edwards P P and Panagopoulos C  1999 
\textit{Phys. Rev. B} \textbf{60} 3244-3251 

\item[] Putilin  S N, Antipov E V, Chmaissen O and Marezio M 1993 
\textit{Nature} \textbf{362} 226-228  

\item[] Ren Y T, Chang H, Xiong Q, Xue Y Y and Chu C W 1994 
\textit{Physica C} \textbf{226} 209-215

\item[] Renker B, Schober H and Gompf F 1996 \textit{J. Low Temp. Phys.} 
\textbf{105} 843-848

\item[] Sacuto A, Combescot R, Bontemps N, Monod P, Viallet V and Colson D
1997 \textit{Europhys. Lett.} \textbf{39} 207-212

\item[] Schilling A, Cantoni M, Guo J D and Ott H R 1993 
\textit{Nature} \textbf{363} 56-58 

\item[] Sette, F, Ruocco G, Krisch M, Masciovecchio C and Verbeni R 1996
\textit{Physica Scripta} \textbf{T66} 48-56

\item[] Stachiotti M G, Peltzer y Blanc\'a E L, Migoni R L, Rodriguez C O, 
Christensen N E 1995 \textit{Physica C} \textbf{243} 207-213 

\item[] Verbeni R, Sette F, Krisch M H, Bergmann U, Gorges B, 
Halcoussis C, Martel K, Masciovecchio C, Ribois J F, Ruocco G and Sinn H 1996
\textit{J. Synch. Rad.} \textbf{3} 62-64

\item[] Yang I S, Lee H G, Hur N H, and Yu J 1995 
\textit{Phys. Rev. B} \textbf{52} 15078-15081

\item[] Zhou X J, Cardona M, Chu C W, Lin Q M, Loureiro S M 
and Marezio M 1996a \textit{Phys. Rev. B} \textbf{54} 6137-6140 

\item[] Zhou X J, Cardona M, Chu C W, Lin Q M, Loureiro S M 
and Marezio M 1996b \textit{Physica C} \textbf{270} 193-206 

\end{harvard}

\end{document}